\documentclass[conference]{IEEEtran}
\IEEEoverridecommandlockouts
\usepackage{cite}
\usepackage{amsmath,amssymb,amsfonts}
\usepackage{blkarray,booktabs,bigstrut}
\usepackage{subfig}
\usepackage{algorithmic}
\usepackage{float}
\usepackage{graphicx}
\usepackage{tabularx}
\usepackage{wrapfig}
\usepackage{pgf-pie}
\usepackage{textcomp}
\usepackage{xcolor}
\usepackage{listings}
\usepackage{multicol}
\usepackage[binary-units]{siunitx}
\usepackage[inline]{enumitem}
\usepackage{multirow}
\usepackage{hyperref}
\newcolumntype{?}{!{\vrule width 1pt}}

\colorlet{punct}{red!60!black}
\definecolor{delim}{RGB}{20,105,176}
\colorlet{numb}{magenta!60!black}

\lstdefinelanguage{json}{
    basicstyle=\normalfont\ttfamily,
    numberstyle=\scriptsize,
    stepnumber=1,
    numbersep=8pt,
    showstringspaces=false,
    breaklines=true,
    literate=
      {:}{{{\color{punct}{:}}}}{1}
      {,}{{{\color{punct}{,}}}}{1}
      {\{}{{{\color{delim}{\{}}}}{1}
      {\}}{{{\color{delim}{\}}}}}{1}
      {[}{{{\color{delim}{[}}}}{1}
      {]}{{{\color{delim}{]}}}}{1},
}

\def\BibTeX{{\rm B\kern-.05em{\sc i\kern-.025em b}\kern-.08em
    T\kern-.1667em\lower.7ex\hbox{E}\kern-.125emX}}
    
\providecommand{\tightlist}{%
  \setlength{\itemsep}{0pt}\setlength{\parskip}{0pt}}
\begin{document}

\title{Monolith Development History for Microservices Identification: a Comparative Analysis}

\author{\IEEEauthorblockN{João Lourenço and António Rito Silva}
\IEEEauthorblockA{INESC-ID, Instituto Superior Técnico, University of Lisbon -- Lisbon, Portugal \\
\{joao.estudante.lourenco, rito.silva\}@tecnico.ulisboa.pt}
}

\maketitle
\thispagestyle{plain}
\pagestyle{plain}
\begin{abstract}
Recent research has proposed different approaches on the automated identification of candidate microservices on monolith systems, which vary on the monolith representation, similarity criteria, and quality metrics used. On the other hand, they are generally limited in the number of codebases and decompositions evaluated, and few comparisons between approaches exist. Considering the emerging trend in software engineering in techniques based on the analysis of codebases’ evolution, we compare a representation based on the monolith code structure, in particular the sequences of accesses to domain entities, with representations based on the monolith development history (file changes and changes authorship). From the analysis on a total of 468k decompositions of 28 codebases, using five quality metrics that evaluate modularity, minimization of the number of transactions per functionality, and reduction of teams and communication, we conclude that the best decompositions on each metric were made by combining data from the sequences of accesses and the development history representations. We also found that the changes authorship representation of codebases with many authors achieves comparable or better results than the sequence of accesses representation of codebases with few authors with respect to minimization of the number of transactions per functionality and the reduction of teams.

\end{abstract}


\section{Introduction}
\label{sec:introduction}

In 2014, Lewis and Fowler~\cite{lewis2014microservices} described microservices, a new architectural style, which was applied at Amazon and Netflix~\cite{netflixamazonmicroservices}. In such systems, instead of a single unit (a "monolith") being responsible for handling all the business logic with a single database, sets of functionalities that implement the logic execute separately in independent services. This brings plenty of advantages, like increased developer productivity, scalability, reliability, maintainability, separation of concerns, and ease of deployment~\cite{alshuqayran_systematic_2016}.
As such, migrating a monolithic application to a microservice architecture is appealing, and different automated approaches have been proposed~\cite{softwarerefactorapproaches,servicecutter,survey2021}. 

These approaches tend to follow a common procedure: (1) collect data; (2) generate a representation of the monolith; (3) define one or more similarity measures for the monolith's elements based on some criteria, and use them to cluster the collected data and generate a decomposition based on the representation and the measures; (4) evaluate the decomposition using one or more quality metrics. However, the techniques at each step vary: as an example, the data collection can be based on the monolith's specifications\cite{servicecutter}, code static analysis~\cite{samuel22}, system execution analysis~\cite{klock2017,hassan17}, development history~\cite{eski_automatic_2018,santos_microservice_2021,lohnertz_steinmetz_2020,mazlami_extraction_2017}, among others. The monolith representation can be based on a graph~\cite{eski_automatic_2018,lohnertz_steinmetz_2020}, a tree~\cite{santos_microservice_2021}, or a sequence of accesses~\cite{samuel22}. Multiple criteria can be used to identify the services of the decomposition~\cite{servicecutter}, like modularity~\cite{eski_automatic_2018}, minimization of the number of distributed transactions per functionality~\cite{nunes19}, or even the reduction of each service's team size~\cite{mazlami_extraction_2017}. The evaluation metrics also vary significantly~\cite{bogner17}.

There has been research on the comparison of the use of code static analysis and system execution analysis data collection techniques~\cite{andrade22}, on development history and lexical monolith's representations~\cite{mazlami_extraction_2017}, and on modularity, team size, and number of transactions criteria~\cite{mazlami_extraction_2017,samuel22}. However, this is limited on the number of possible combinations, and so, more research has to be done. In particular, due to the emerging trend in software engineering in techniques based on the analysis of codebases' development history~\cite{kalliamvakou16}, it became especially interesting to compare approaches that use the monolith code structure, in particular the sequences of accesses to domain entities, which is one the monolith representations more often used, e.g.~\cite{jin21}, with representations based on the monolith development history. The latter simplify the collection step, because they are independent of the programming language and technology used in the monolith implementation.

Therefore, our research question is: \textbf{How do monolith microservices identification approaches that use the monolith development history based representations perform when compared with approaches that use the monolith functionalities sequences of accesses representation?}

The research question is evaluated according to the criteria of modularity, minimization of the number of transactions per functionality, reduction of teams and communication, whereas metrics are used for the assessment. A total of 28 monolith codebases are used for the empirical study.

The results show that in most cases, approaches based on the sequence of accesses monolith representation (used for the transactions minimization criteria) perform better for complexity and coupling metrics, but worse for the cohesion metric, than those using a monolith development history based representation. This is also verified for a combination of both representations. However, the vast majority of the decompositions with the best values in these quality metrics are made with a combination of both representations. This means that when the right weights for the different similarity measures are found, it is possible to obtain better decompositions when both types of representation are combined. With regards to the team size reduction ratio, we also found that the sequence of accesses representation with the transaction minimization criteria achieves only slightly worse results than the development history based representation with team size or modularity criteria.

This section presented the context of our work. Section~\ref{sec:background} details the measures and metrics we will use to evaluate our solution; Section~\ref{sec:related-work} describes different existing approaches to automated migration, and what work has been done on the comparison of approaches; Section \ref{sec:implementation} outlines our implementation and various choices that had to be made; Section~\ref{sec:evaluation} contains comparisons between the previous results and our own; Finally, section~\ref{sec:conclusion} presents final thoughts and a summary of the paper.

\section{Monolith models and decomposition techniques}
\label{sec:background}

We leverage on previous work to compare the use of the different monolith representation models for the identification of microservices in monolith systems: (1) the sequence of accesses~\cite{samuel22}, and (2) development history~\cite{mazlami_extraction_2017}. While the former representation is one of the most widely used by different approaches, the latter seems promising in terms of the new trends on code repositories mining and the relative independence of programming languages and software frameworks used.

The use of sequences of accesses to represent a monolith requires the identification of its set of functionalities, $F$, and the accessed domain entities, $E$. For each functionality $f \in F$ there is a callgraph, $f.graph$ that captures the sequences of accesses done to the domain entities. Each access, $a \in A$, is a pair $(e,m)$, where $e \in E$ and $m \in \{r, w\}$ ($r$ is the read access and $w$ the write one). Given an entity $e \in E$, $e.funct(m)$ denotes the set of functionalities that have an access in the entity according to access mode $m$; if $m$ is omitted, then it can be any type of access. Note that this representation can be obtained either through a static analysis or a system execution analysis of the monolith.

On the other hand, the use of the development history monolith representation requires the identification of its set of commits, $C$, where each commit, $c \in C$, contains the set of files, $c.files \in F$ that where changed together, where $F$ represents the set of all files in the codebase. Additionally, a commit $c \in C$ author, $c.author \in A$, belongs to the set $A$ of codebase developers. Given a file $f \in F$, $f.authors$ denote the set of authors that have a commit in the file. Finally, a commit $c \in C$ also contains the time when it occurred, $c.time$.

A decomposition is a partition of the monolith domain entities. A decomposition $d \in D$, where $D$ represents the set of all decompositions, is a set of clusters, $d.clusters \in 2^E$, of the monolith domain entities. Therefore, $\forall_{cl_i,cl_j \in d.clusters} cl_i \cap cl_j = \emptyset$ and $\bigcup_{cl \in d.clusters} cl = E$. Note that for decompositions generated using the development history representation, it is necessary, in some part of the processing pipeline, to filter the files that correspond to the domain entities.

A decomposition generation is driven by similarity measures between representation elements, which can be domain entities or files. The smaller the distance between them, the higher is the likelihood they belong to the same microservice.

The similarity measures for the sequence of accesses representation are defined over the relation between the functionalities and the entities they access, such that the number of distributed transactions per functionality is minimized:

\begin{itemize}
    \item Access measure - The access similarity of two entities $e_i, e_j \in E$ depends on the likelihood of functionalities accessing (by reading or writing) both $e_i$ and $e_j$, such that a functionality can completely execute in a single cluster:
    \begin{equation}
        sm_{access}(e_i,e_j) = \frac{\#(e_i.funct \cap e_j.funct)}{\#e_i.funct}
    \end{equation}
    \item Read measure - The read similarity of $e_i, e_j \in E$ depends on the likelihood of functionalities reading both $e_i$ and $e_j$, such the reads of a functionality are done in a single cluster:
    \begin{equation}
        sm_{read}(e_i,e_j) = \frac{\#(e_i.funct(r) \cap e_j.funct(r))}{\#e_i.funct(r)}
    \end{equation}
    \item Write measure - The write similarity of $e_i, e_j \in E$ depends on the likelihood of functionalities writing both $e_i$ and $e_j$, such that the writes of a functionality occur in a single cluster:
    \begin{equation}
        sm_{write}(e_i,e_j) = \frac{\#(e_i.funct(w) \cap e_j.funct(w))}{\#e_i.funct(w)}
    \end{equation}
    \item Sequence measure - The sequence similarity of $e_i, e_j \in E$ depends on the number of consecutive accesses to both entities across all functionalities, such that the number inter-cluster invocations is minimized:
    \begin{equation}
        sm_{sequence}(e_i,e_j) = \frac{sumPairs(e_i,e_j)}{maxPairs}
    \end{equation}
    where $sumPairs$ represents number the consecutive occurrence of accesses to the entities, and $maxPairs$ the maximum of all sums of pairs.
\end{itemize}

For the development history based representations the similarity measures are defined on the co-occurrences of changes between two files, and the common authorship of files changes:

\begin{itemize}
    \item Commits - where files that were changed together more often are likely to stay in the same cluster:
    \begin{equation}
    \label{eq:commit-similarity}
    sm_{commit}(f_i, f_j) = \frac{\# \{c \in C: f_i,f_j \in c.files \}}{\#\{c \in C: f_i \in c.files\}}
    \end{equation}

    \item Authorship - where files that were changed by the same developers are likely to stay in the same cluster:
    \begin{equation}
    \label{eq:author-similarity}
    sm_{author}(f_i, f_j) = \frac{\#(f_i.authors \cap f_j.authors)}{\#f_i.authors}
    \end{equation}

\end{itemize}

To evaluate a certain decomposition several quality metrics are defined, each focusing on a particular quality of the decomposition like modularity, migration effort or team reduction size. The \textbf{complexity} of the decomposition is informally defined as the effort required to perform the decomposition, due to the intermediate states that are naturally introduced, as described by the concept of Saga~\cite{Garcia-Molina87,richardson2018microservices} for the implementation of distributed transactions with eventual consistency~\cite{Shapiro09}. The value of the complexity metric~\cite{rito_complexity_2020} increases with the number of intermediate states created by the distributed transactions, because they have to be considered in the implementation of the functionalities business logic, due to the lack of isolation. The \textbf{uniform complexity} is based on the complexity, and is computed by dividing the complexity of a decomposition by the maximum possible complexity, achieved when each entity is in a single cluster and only accesses are considered.
The \textbf{cohesion} indicates the percentage of entities of a microservice accessed whenever there is an access, and so it varies between 0 and 1. If the cohesion of a decomposition is 1, then whenever any microservice is accessed, all of its entities are accessed, so it strongly follows the Single Responsibility Principle~\cite{Martin2005}. \textbf{Coupling} is a metric applied to two microservices, and describes the percentage of entities that a service exposes to the other. The coupling of a decomposition corresponds to the average coupling among all pairs of microservices. A coupling of 1 means that all services expose entities to all other services, which is a very high inter-service dependency and an undesirable trait. The \textbf{team size reduction}, or \textit{tsr} for short, indicates if the average team size is shorter, by comparing the average number of authors per microservice to the total number of authors ~\cite{mazlami_extraction_2017}. A \textit{tsr} of 1 would indicate no reduction in team size, as all services have the same number of authors that the original monolith had. Finally, a \textbf{combined} metric is used to sum up the results of all other metrics. It is a number between 0 and 1, where 0 is a perfect
decomposition in all metrics, and 1 is the worst decomposition
possible in all metrics.

\section{Related Work}
\label{sec:related-work}

Different approaches exist to decompose a monolith system into a service-oriented architecture, with various input sources, final service granularity, and applicabilities. A recent overview was done in~\cite{softwarerefactorapproaches}, where four main categories are highlighted: 

\textit{Meta-data aided approaches}. These make use of various representations of code, and not the code itself, to suggest decompositions. In~\cite{Ahmadvand2016}, it is suggested a manual conceptual approach that makes use of use-cases, security, and scalability requirements to partition a monolith. The authors in~\cite{Baresi2017} make semantic evaluations on a monolith's OpenAPI specification, and suggest a decomposition of interfaces based on the similarity of terms in the specification's descriptions. They then evaluate the decompositions based on granularity, cohesion, and coupling. Service Cutter, presented in~\cite{servicecutter}, is a very thorough tool making use of 16 different coupling criteria to drive decompositions. Different criteria require different inputs, which could be use cases, entity relationships models described in UML, the entities themselves, and so on. Finally, in~\cite{mazlami_extraction_2017}, they extract information from a monolith's development history (stored in a version control system), and extract logical, authors, and lexical relationships between classes.

\textit{Static code analysis aided approaches}. The focus here is on the code itself, from which a representation is generated and a decomposition is made. In~\cite{Escobar2016}, they parse the annotations in classes of Java Enterprise Edition applications to identify domain entities and their types. They are then organized into an Abstract Syntax Tree according to their relationships, and a clustering algorithm extracts microservices suggestions. In ~\cite{Levcovitz2016}, the authors extract the dependencies between business functions, databases, and facades, and build a dependency graph based on the relationships between these elements. The microservices suggestions are done through manual code inspection. In both cases, entities play a central role in the whole process.

\textit{Workload-data aided approaches}. The execution process is analysed to extract relationships, and then a decomposition is derived. Only one approach is described in this category~\cite{Mustafa2017}, where the execution logs of a web application are parsed and pages that have higher workloads are candidates to be split into microservices.

\textit{Dynamic microservice composition approaches}. These approaches keep generating a decomposition until achieving a stable and desirable state. In~\cite{hassan17}, the authors reason that it's not possible to fully capture the expected behaviour and granularity of microservices at design time. Therefore, they create architectural elements with variable boundaries, that get changed according to the behaviour when executing the monolith. In~\cite{klock2017}, the authors extract features (chunks of functionalities that implement some business logic) and register their usage and performance during the execution of the system. A genetic algorithm reorganizes features in different microservices according to the observed usage and performance, so that the cohesion is maximized and the coupling is minimized.

In short, the approaches suggest the usage of different information sources to represent the monolith, depending on whether they focus on its structure~\cite{Ahmadvand2016,Baresi2017,servicecutter,Escobar2016,Levcovitz2016}, behaviour~\cite{Mustafa2017,hassan17,klock2017}, or development process~\cite{mazlami_extraction_2017}. Additionally, different decomposition criteria are used, like modularity~\cite{Ahmadvand2016,Baresi2017,servicecutter,mazlami_extraction_2017,Escobar2016, Levcovitz2016,hassan17}, performance~\cite{Mustafa2017, klock2017}, team size~\cite{mazlami_extraction_2017}, and security/usage requirements~\cite{Ahmadvand2016,servicecutter}.

Some research has been done on the comparison of different data collection and decomposition strategies. In~\cite{andrade22}, the authors compare which of two data collection approaches, static code analysis or dynamic analysis, generate better decompositions when considering a criteria of reducing the number of distributed transactions. An evaluation was performed on two systems, and it was found that no approach outperforms the other, but the dynamic analysis required more effort. In~\cite{samuel22}, sequences of accesses to domain entities were statically collected, and four different similarity measures were defined, all driven by the goal to reduce the number of distributed transactions. The measures covered different types of accesses, like writes, reads, writes or reads, and sequences. By analysing decompositions from 121 codebases, it was found that there isn't a single measure or combination of measures that yields better results, in terms of the migration's complexity. In~\cite{mazlami_extraction_2017}, information about the authors that changed each file and which files change together most often was collected from the monolith's development history. Lexical data was collected from the monolith's latest source code. The similarity measures used were logical coupling, contributor coupling, and semantic coupling, and the generated decompositions and were evaluated on how well the original monolith's author set was partitioned in the microservices (team size reduction ratio) and how much domain redundancy was obtained. It was found that any combination of the three measures had good results, with the contributor coupling displaying more dispersion.

There is still a lack of understanding on how the different approaches compare in terms of the results they produce. Our work focuses on comparing the sequence of accesses and development history based representations, using
four similarity measures from~\cite{samuel22}, and two measures similar to the logical coupling and contributor coupling from~\cite{mazlami_extraction_2017}.

\section{Implementation}
\label{sec:implementation}

The sequences of accesses monolith representation, that we leverage to perform comparisons, is described in depth in~\cite{samuel22}. This representation is obtained by using Spoon, a Java source code analyser. They identify controllers and domain entities in Spring-Boot monoliths, and parse method calls in each controller to obtain the types of accesses made to the domain entities. From this analysis results a JSON file containing the controllers (functionalities) as keys, and the accesses to entities that the controller can perform as values.

On the other hand, the development history based representation is obtained by parsing the output of the \texttt{git log} shell command with Python, and applying several processing techniques to ensure that data related to renamed or deleted files is accurate. The representation consists of two JSON files: one with the monolith's files as keys and the number of times each file changed with others as values, named file changes representation; and another with the files as keys and the authors that changed each file as values, named changes authorship representation. The combination of both of these representations is a development history based representation.

Then, we compute the similarity based on the monolith's representations - similarity between domain entities for the sequences of accesses and between files for development history. These are used by a hierarchical clustering algorithm to generate the decomposition. 

Our analysis tool then computes the complexity, cohesion, coupling, and team size reduction ratio for a large number of decompositions, obtained by varying the weights and the number of clusters.

\begin{figure}
    \centering
    \includegraphics[width=0.48\textwidth]{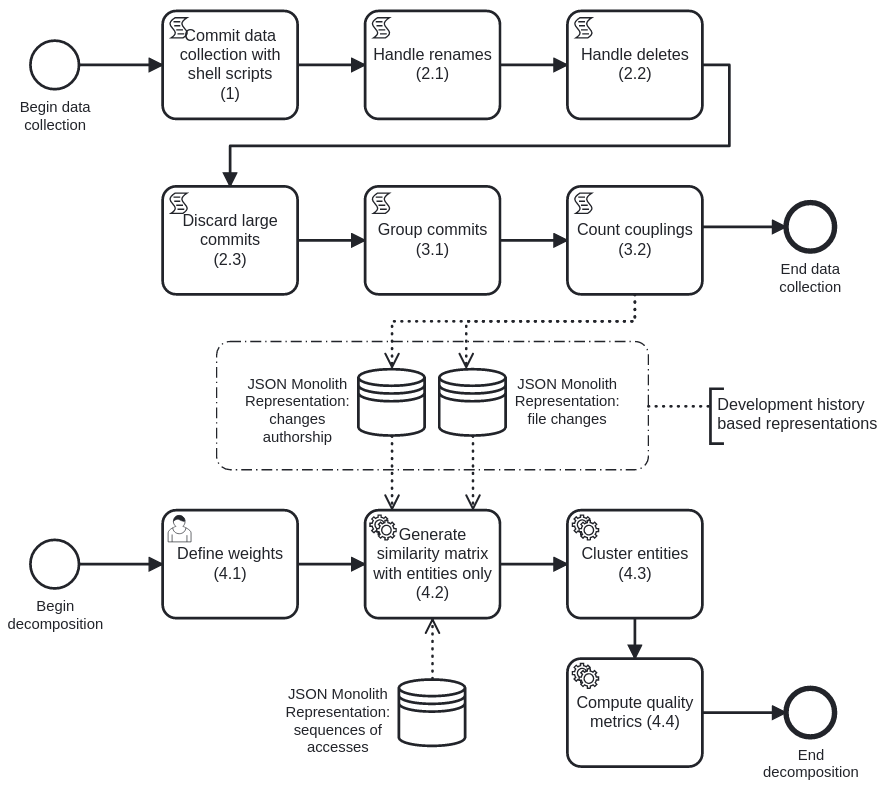}
    \caption{Pipeline for generating a decomposition of a single monolith system, using data extracted from the development history and data from sequences of accesses. The numbering on the different scripts/processes is referenced throughout the section.}
    \label{fig:history-pipeline}
\end{figure}

Figure \ref{fig:history-pipeline} contains the overview of the processing pipeline necessary to create a decomposition of a monolith. It includes a detailed view of the data collection steps, and a higher-level overview of the decomposition steps. Note that for the generation of the similarity matrix (step 4.2), the data from the sequences of accesses is also used. This data is obtained from a separate data collection process.

\subsection{Initial Data Collection}

The very first step is the data collection (1), which be achieved in three sub-steps: a \textit{git log} command, \textit{awk} processing, and \textit{grep} filtering.

\begin{figure}
    \centering
    \includegraphics[width=0.48\textwidth]{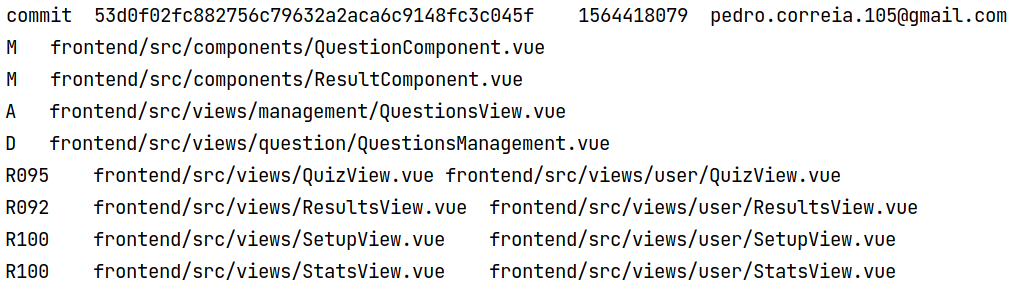}
    \caption{Example of the output from the commit data extraction script.}
    \label{fig:commit-extraction-data-qt}
\end{figure}

The \textit{git log} shell command used is the following:

\begin{lstlisting}[language=bash]
git log
--reverse
--name-status
--find-renames
--pretty=format:"commit	%H %ct %ce"
\end{lstlisting}

This tells Git to return all commits in chronological order, by displaying the names and statuses of changed files, detecting and reporting renames of files, and displaying the literal string "commit", followed by the commit hash (\verb+%H+), the commit time formatted as the UNIX timestamp (\verb+%ct+), and the commit author's e-mail (\verb+%ce+). Figure~\ref{fig:commit-extraction-data-qt} contains a partial output of the script applied to a
codebase. On the first line, you can see the string "commit", followed by the hash, the commit time, and the author's e-mail. This is followed by one line for each of the files changed in this particular commit. The line contains the status, which can be \textbf{A} (Added), \textbf{D} (Deleted), \textbf{M} (Modified), or \textbf{R} (Rename), as well as the file's name. \textbf{R} is followed by a number between 50 and 100 that represents the similarity between the file in the last commit and the file in the current commit, and the line also contains the new name.

The output from the git log command is stored in a file, and some \textit{awk} lines take care of converting it to a csv-like format. This way, we can build a script in Python to read the data into a dataframe for further processing. The dataframe is a data structure that can be seen as a very efficient table, with rows and columns, and is provided by the Pandas package. We use it extensively in our implementation, as we manage to hit up to 10x faster data processing speeds compared to native Python lists and dictionaries.

A final \textit{grep} command filters out any file that does not end in \texttt{.java}.  This is easily configured, making the data collection part independent of the language used in the monolith. 

\subsection{Data Cleaning}

At this point, the data from the logs is not yet ready to be used. There are two situations regarding files getting renamed and files getting deleted that require special care. 

If a file is renamed from \verb+A.java+ to \verb+B.java+, and no care is taken, the final decomposition would contain both files because there have been changes to both files (before and after the rename) in the history. This is incorrect because \verb+A.java+ no longer exists. But since it is the same file as \verb+B.java+, we replace all instances of \verb+A.java+ with \verb+B.java+ (step 2.1). A similar strategy is followed by Mazlami in~\cite{mazlami_extraction_2017}.

Files may be deleted at timestamp $X$ but then appear as added or modified in timestamp $X + Y$, either by getting merged from another branch or by being re-added. If this happens, we don't want to delete those files: there is relevant information after their supposed deletion, and they still exist in the current snapshot of the repo. So, we delete any occurences of files that show up at least once with a \verb+DELETED+ status, but \textit{only if} they don't show up at a later timestamp with a different status (step 2.2).

Finally, just like in~\cite{ying_predicting_2004}, we discard any commits with more than 100 modified files for the purposes of data analysis - but \textit{not} for the purpose of identifying renames and deletes. The reasoning is that these types of commits are usually associated with refactor operations, where a large batch of files is renamed, deleted, or otherwise changed. But if we were to completely ignore these commits when identifying renames and deletes, we would be throwing out key moments in history that help us in this task. 

\begin{figure}
    \centering
    \includegraphics[width=0.48\textwidth]{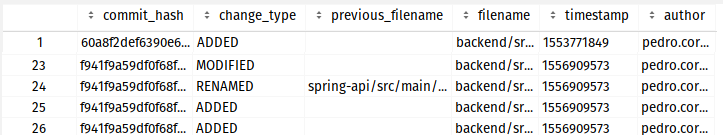}
    \caption{The head of the history dataframe of the codebase.}
    \label{fig:history-dataframe}
\end{figure}

At the end of this process, we have a dataframe containing one row for each file change. The first few rows of the 
codebase can be found in figure \ref{fig:history-dataframe}. For every file that was Added, Modified, or Renamed, there's a row in the dataframe containing the commit hash where it appeared, the type of the change, the previous filename (if the change is a Rename), the filename, the timestamp of the change and the change's author. With this, we have every information we need to compute the similarities between files/entities.

\subsection{Coupling Computation}
The goal is to convert the history dataframe obtained before into a JSON structure that states which files have been modified together, and another JSON that states which authors have modified a given file.

Grouping commits is the first step towards this goal. Different works mention bundling together commits performed by the same author in a short timeframe~\cite{ying_predicting_2004,kagdi_mining_2007,mazlami_extraction_2017}. The reasoning for this is that it's likely that all those commits relate to the same task, so they should be considered as just one commit. We opted to also follow this reasoning, and consider all sequential commits made by the same author in the period of one hour as just one commit. Then, for each commit, we generate permutations of size 2 of all files that were modified in that commit. For each of these pairs, a JSON structure is updated with new coupling counts, and another JSON structure is updated with authors.

\begin{figure}
    \centering
    \includegraphics[width=\linewidth]{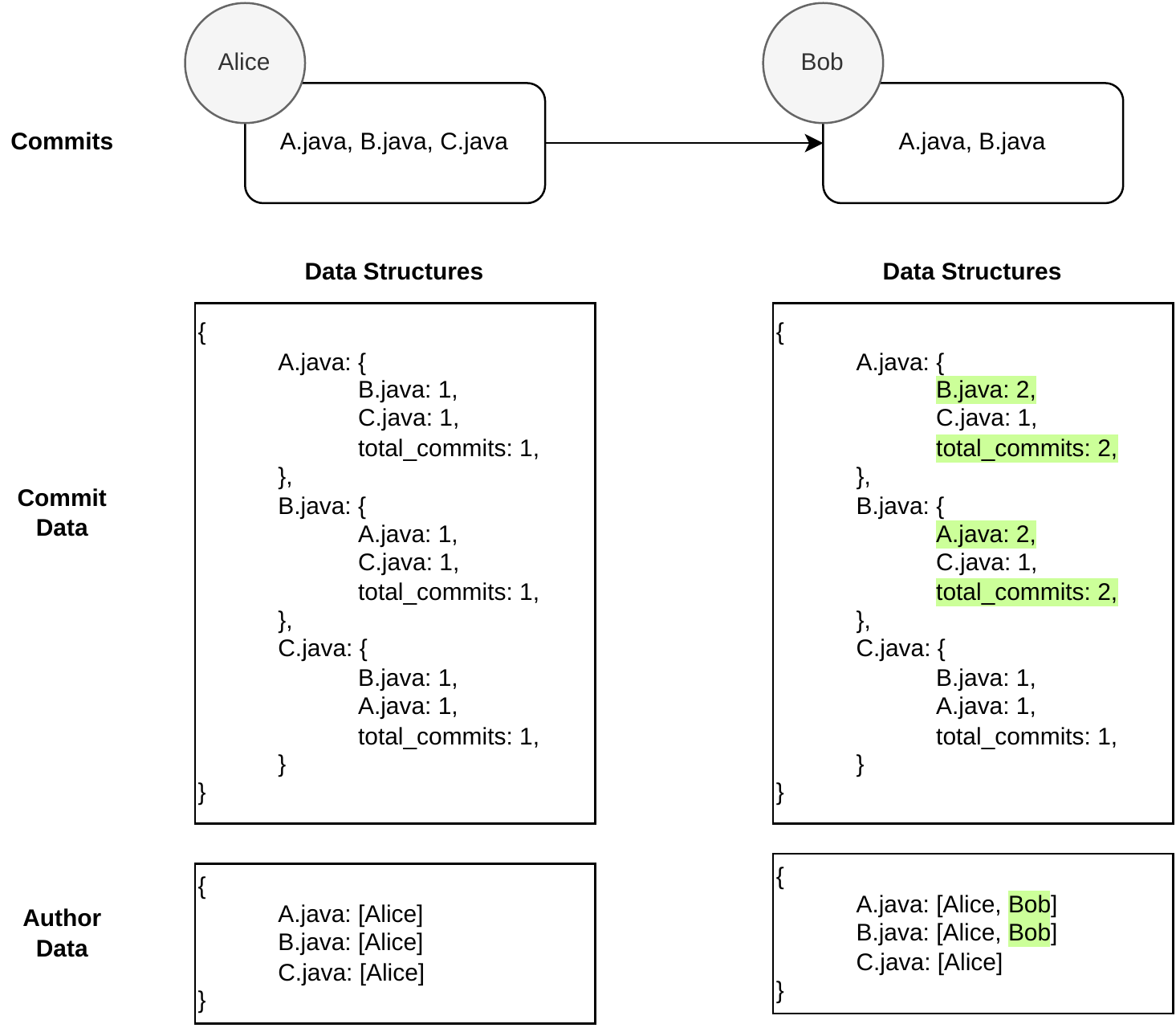}
    \caption{How the structures containing the commit and author data look like, and how they change when a new commit is parsed. The change from the first to the second commit is highlighted in green.}
    \label{fig:commit-coupling-structures}
\end{figure}

Figure~\ref{fig:commit-coupling-structures} illustrates how this structure is updated and how it looks like in a simple situation with two commits. For each file found in the file changes representation, we can very simply query how many times it changed with other files, as well as the number of commits it appeared in.

\subsection{Performing a decomposition}
Decomposing a monolith into microservices is done with an automated tool and follows a similar procedure as to what is described in the literature, applying the similarity measures in equations 1-6. In what concerns the similarity measures in equations 5 and 6, only the files corresponding to domain entities are automatically selected from the monolith’s data representations.

The similarity matrix contains one line and one column for each domain entity, and each entry in the matrix has a number that represents how similar the two entities are, and combines both the file changes information and the changes authorship information according to the defined weights. We can choose a value between 0 and 100 for the weight, as long as the sum of weights equals 100. 

We apply a hierarchical clustering algorithm to this matrix, and perform a cut with a certain number $N$. This results in a decomposition of the monolith's domain entities into $N$ microservices, each one containing at least one entity. Pairs of entities that have higher values in the matrix will generally be placed in the same microservice.

With this strategy, it's easy to perform decompositions that also take into account data from the sequences of accesses. Since both our measures and these static measures vary between 0 and 1, we can perform a pondered sum of their values, and the final similarity of any two entities will also vary between 0 and 1. The ability to do this serves as the basis for the evaluation and comparisons made in Section \ref{sec:evaluation}.

\subsection{Evaluating a decomposition}

For each decomposition, four metrics are computed: complexity, cohesion, coupling, and team size reduction. The first three rely on the data from the monolith sequences of accesses representation, because they are based on the accesses to domain entities. This means that this representation must always be available even when performing a decomposition only with the development history representation. The team size reduction metric depends only on how the entities were split into the various services and the authors that changed the entities. This means that the history representation must be available when computing this metric for decompositions made only with the sequence of accesses representation.

We also compute a combination of all these metrics, for a decomposition $d \in D$, in the following fashion:

\begin{equation}
    comb(d) = \frac{comp(d) + coup(d) + tsr(d) - coh(d) + 1}{4}
\end{equation}

This yields a number between 0 and 1, where 0 is a perfect decomposition in all metrics, and 1 is the worst decomposition possible in all metrics.

\section{Evaluation}
\label{sec:evaluation}

\hypertarget{approach}{%
\subsection{Approach}\label{approach}}

The research goal, as stated in the introduction, is the
following:

\textbf{How do monolith microservices identification approaches that use the monolith code evolution representation perform when compared with approaches that use the monolith functionalities sequences of accesses representation?}

To address this research question, we generate a large set of
decompositions and analyse how the decomposition qualities vary
depending on different monolith representations and similarity measures.

Whenever we create a decomposition, we choose the weights to attribute
to our similarity measures. If we choose 0 as the weight for the commit
and the author similarity, then the decomposition is created using only
data from the sequences of accesses representation. On the other hand,
if we choose 0 as the weight for all four sequences of accesses measures
(access, read, write, sequence), then the decomposition is created using
only data from the development history based representations. In the
remaining combinations, the decompositions are created with data from
both sources. By filtering the results of all generated decompositions,
we can obtain five distinct groups of decompositions based on the
weights used to create them: only data from the file changes
representation; only data from the changes authorship representation;
only data from the sequences of accesses representation; only data from
the file changes representation and changes authorship; data from all
representations. This allows us to then compare the groups' quality
metrics and draw conclusions.

The comparisons will first be made by evaluating the median and the
dispersion of each quality metric in each of these five groups, which
gives us an overview of how, on average, each representation behaves. We
also evaluate the median values of the metrics in the case of the best
decompositions - that is, the decomposition of each codebase with the
best value for each metric. This gives us a different perspective by
focusing on which group performs best when the ideal weights for the
similarity measures are found. Finally, we assess if our findings hold
when we compare codebases with more commits and authors than the mean
with codebases with less commits and authors than the mean.

The comparisons are, initially, done visually through the analysis of
boxplots. Whenever it's not visually obvious that there is a large
difference between groups, and considering that there are different
amounts of decompositions in each group and the quality metric values
don't follow a normal distribution, we use a Welch T-test with the
following hypotheses:

\begin{itemize}
\tightlist
\item
  \(H_0\) - There are no significant differences between the mean
  \$\{QUALITY METRIC\} of \$\{GROUP\} decompositions and \$\{OTHER
  GROUP\}.
\item
  \(H_1\) - The mean \$\{QUALITY METRIC\} of \$\{GROUP\} decompositions
  is greater than \$\{OTHER GROUP\}.
\end{itemize}

\hypertarget{codebases-and-decompositions-characterization}{%
\subsection{Codebases and decompositions
characterization}\label{codebases-and-decompositions-characterization}}

In \cite{samuel22}, 121 codebases were chosen for evaluation. The
codebases were selected from GitHub, using the following procedure:

\begin{enumerate}
\def\labelenumi{\arabic{enumi}.}
\tightlist
\item
  Get all GitHub repositories that list the Spring Data JPA library as a
  dependency;
\item
  Filter out repositories that did not contain at least 5 files whose
  name ended in \texttt{Controller.java}, and at least 5 files whose
  name did not contain \texttt{Dao} or \texttt{Repository}.
\item
  The remaining repositories were ordered by the number of GitHub stars,
  and 118 codebases were manually selected. Repositories from lessons or
  tutorials, and repositories that did not use just Spring Data JPA were
  disregarded.
\end{enumerate}

From this list, we selected 28 codebases by choosing those with at least
100 commits, and at least 2 authors. The codebases under study have a
mean commit count of 3080.8 with a standard deviation of 7604.5 (median
of 472), and a mean author count of 18.7, with a standard deviation of
30.1 (median of 6.5). This shows that we are only considering projects
with an actual development history, and not toy-like or quickly
abandoned projects.





For each codebase, we create decompositions with 3 to 10 clusters,
according to the number of entities it contains:
\(3 \leq n\_entities < 10 =\) 3 clusters; \(10 < n\_entities < 20 =\) 3,
4, and 5 clusters; \(n\_entities \geq 20 =\) 3 to 10 clusters. For each
number of clusters, we generate decompositions with varying weights on
the six measures: from 0 to 100, with increments of 10. The distribution
of the number of generated decompositions can be found in Table
\ref{tab:decompositions-characteristics}.

\begin{table}[H]

\caption{\label{tab:decompositions-characteristics}The number of generated decompositions across all codebases.}
\centering
\begin{tabular}[t]{llrr}
\toprule
\#Entities & \#Clusters & \#Codebases & \#Decompositions\\
\midrule
3 to 9 & 3 & 4 & 11982\\
10 to 19 & 3 to 5 & 8 & 72073\\
20+ & 3 to 10 & 16 & 384384\\
\midrule
\textbf{Total} & \textbf{} & \textbf{28} & \textbf{468439}\\
\bottomrule
\end{tabular}
\end{table}

\hypertarget{comparison-of-representations}{%
\subsection{Comparison of
representations}\label{comparison-of-representations}}

When evaluating the quality metrics of all decompositions of each
representation, we found that the development history representation
showed better results when it comes to cohesion than the sequence of
accesses, and the sequence of accesses, despite containing no data
regarding changes authorship, still performs well with regards to the
\emph{tsr}. For the coupling, complexity, and combined metrics, the
sequence of accesses representation often generated better
decompositions than the remaining representations.

Having concluded this, we now turn our analysis to the best
decompositions of each codebase. If we group all decompositions by
codebase and by number of clusters, and identify the minimum uniform
complexity decompositions in each group, we find that 92.95\(\%\) of the
best decompositions in terms of complexity were made by combining data
from the development history representations with the sequences of
accesses. Furthermore, 80.77\(\%\) of the best decompositions in terms
of cohesion, 95.51\(\%\) of the best decompositions in terms of
coupling, 95.51\(\%\) of the best decompositions in terms of \emph{tsr}
and 91.67\(\%\) of the best decompositions in terms of combined were
also obtained by this combination of representations. This finding,
together with the overview of the previous subsections, suggests that
the vast majority of weights combinations yield worse results than
considering just data from the sequences of accesses; but, if the right
combination is found, the results can be improved in almost all cases.

\begin{figure*}

{\centering \subfloat[Uniform complexity\label{fig:best-decompositions-compare-1}]{\includegraphics[width=0.3\linewidth]{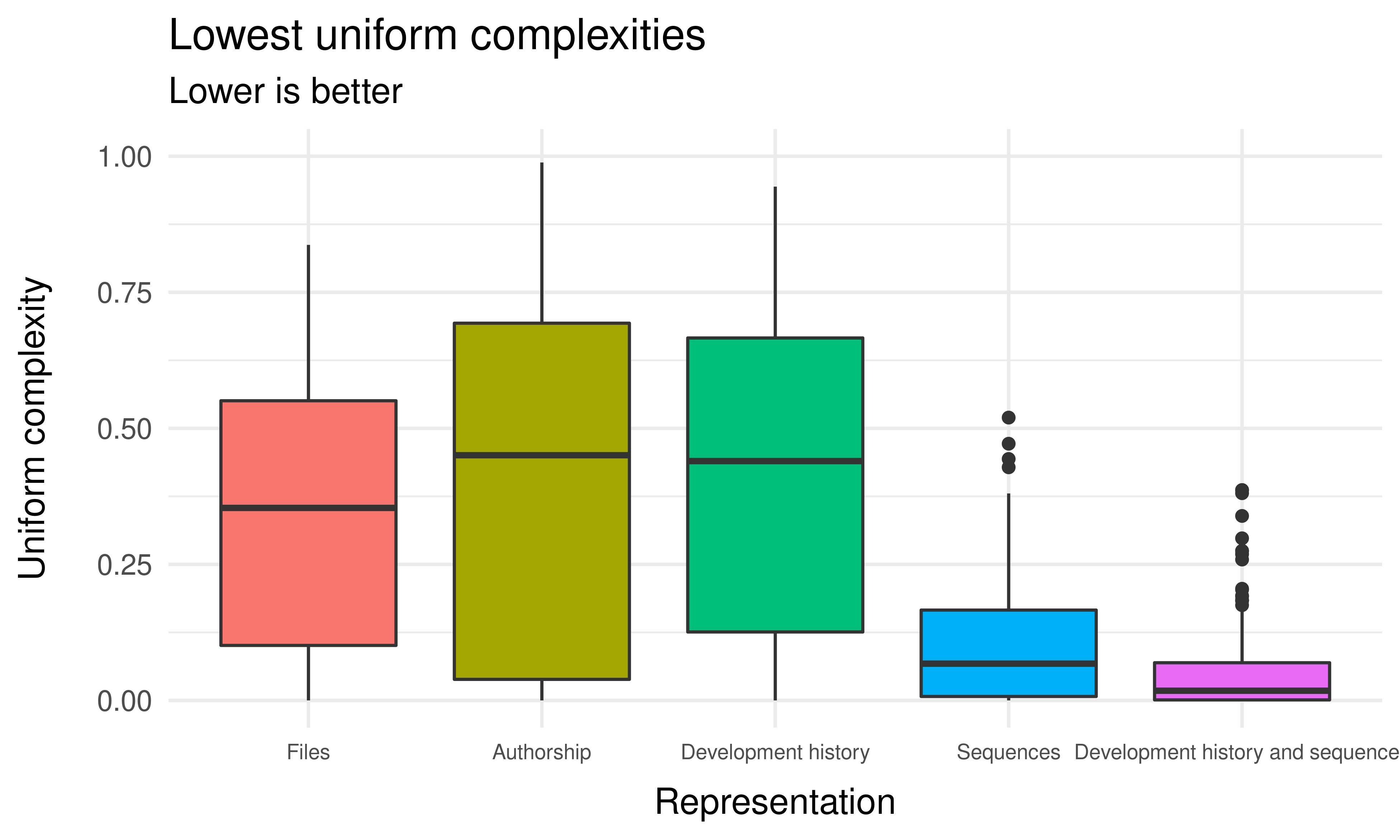} }\subfloat[Cohesion\label{fig:best-decompositions-compare-2}]{\includegraphics[width=0.3\linewidth]{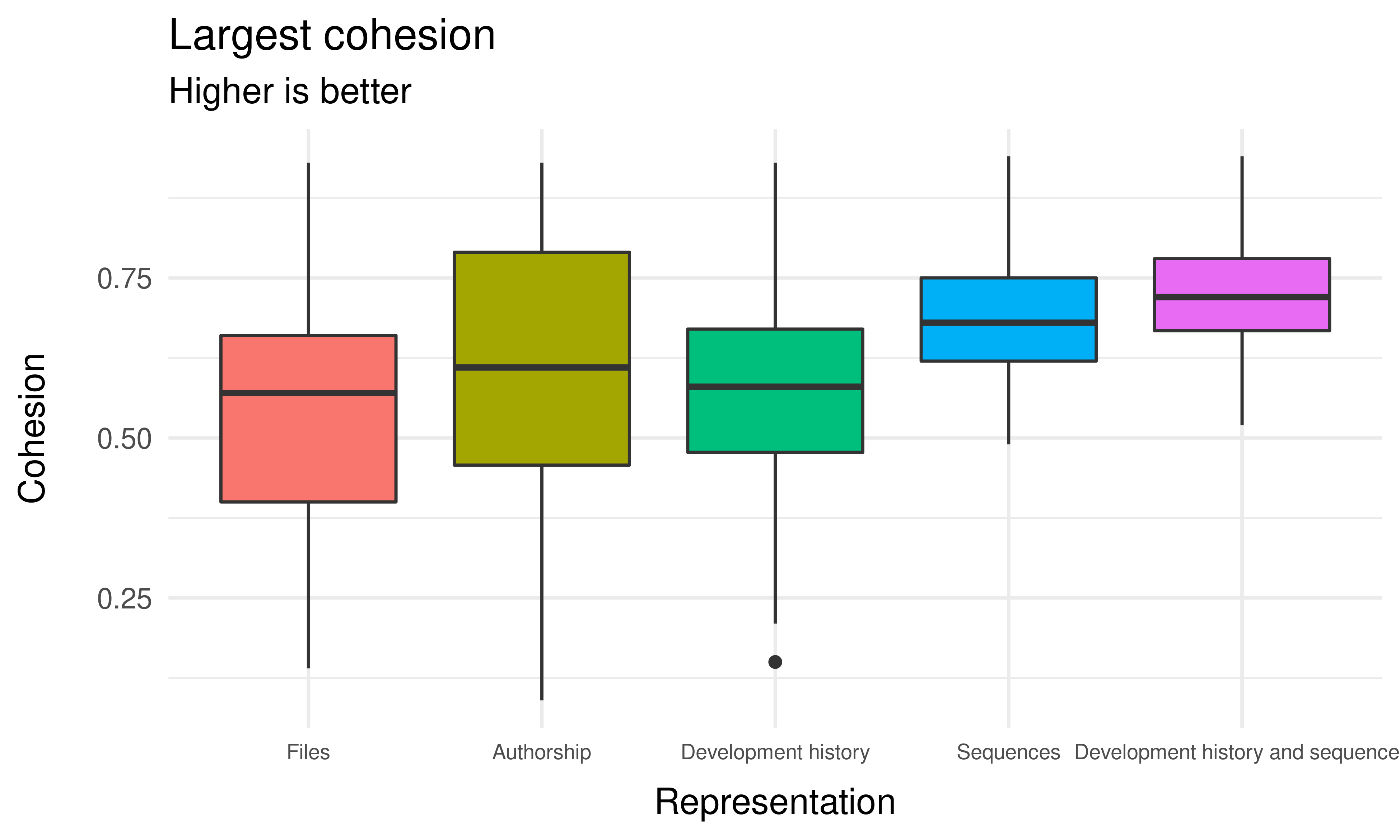} }\subfloat[Coupling\label{fig:best-decompositions-compare-3}]{\includegraphics[width=0.3\linewidth]{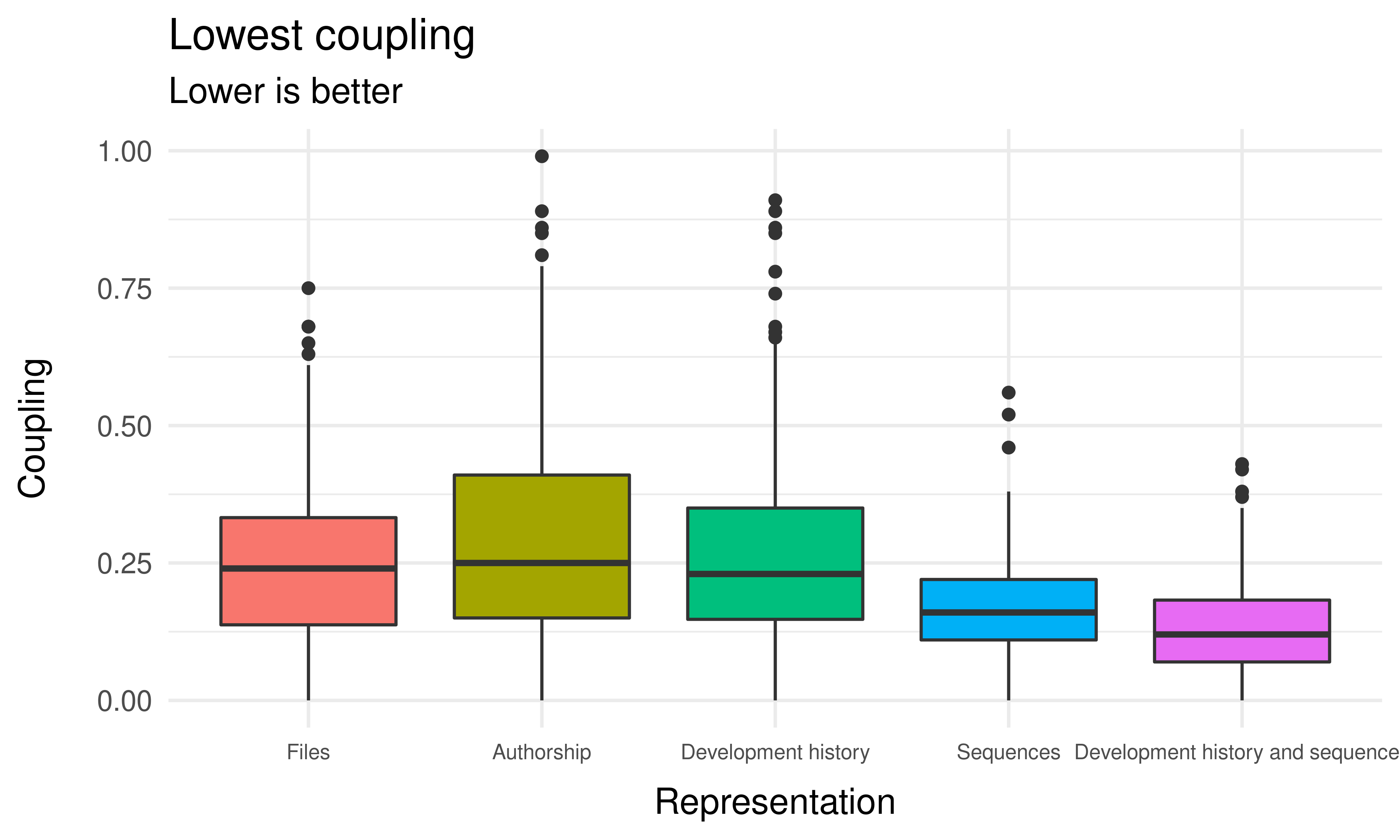} }\newline\subfloat[Team size reduction ratio\label{fig:best-decompositions-compare-4}]{\includegraphics[width=0.3\linewidth]{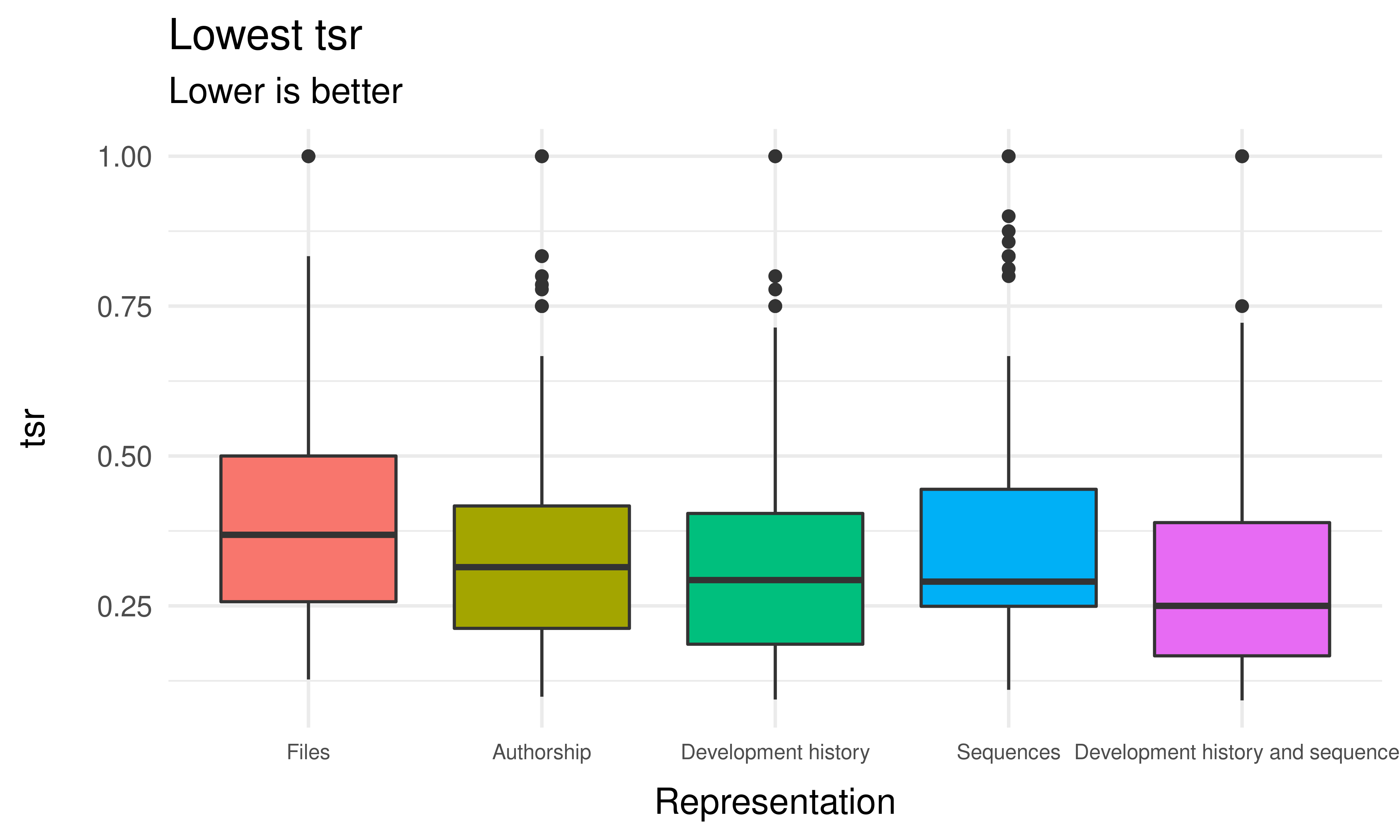} }\subfloat[Performance\label{fig:best-decompositions-compare-5}]{\includegraphics[width=0.3\linewidth]{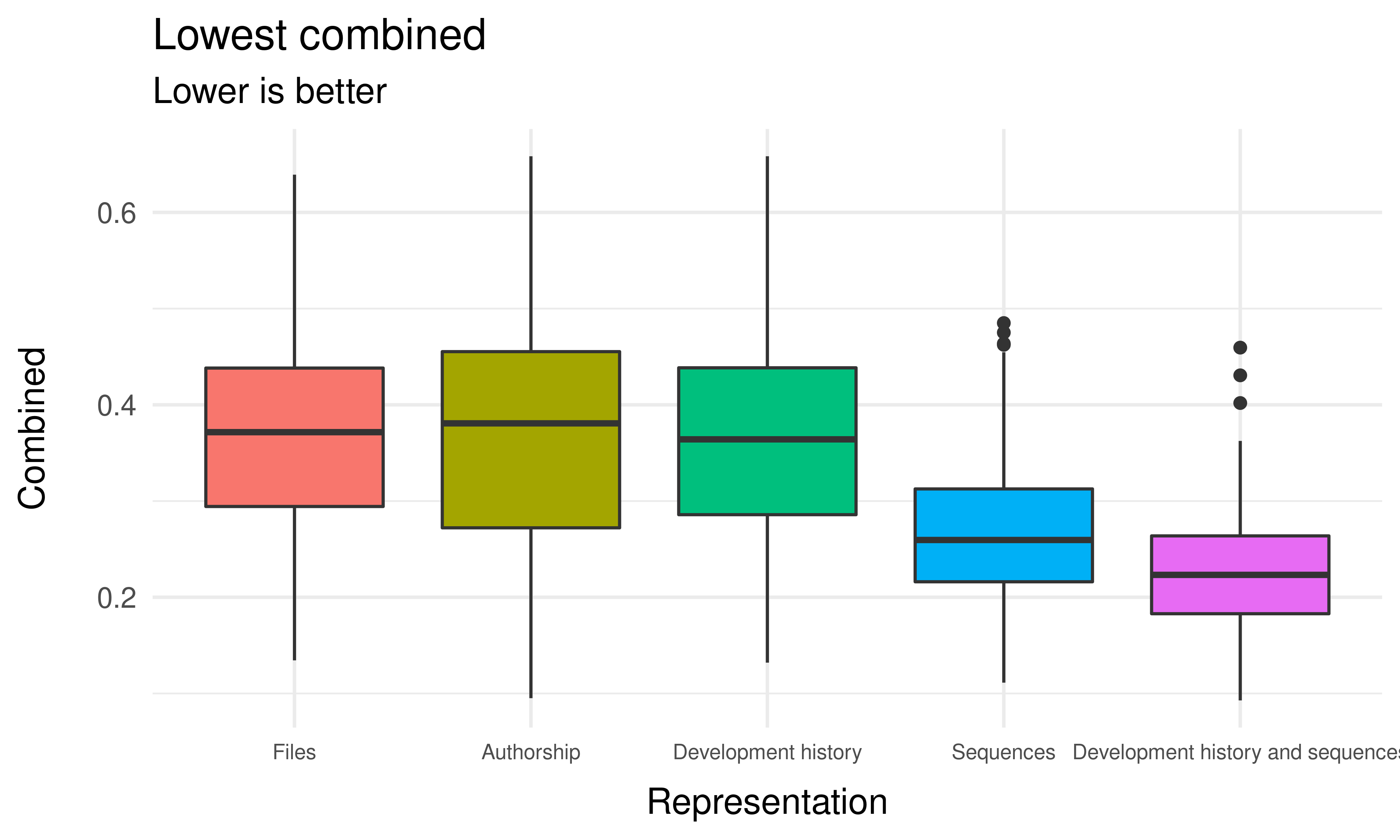} }\subfloat[Legend\label{fig:best-decompositions-compare-6}]{\includegraphics[width=0.3\linewidth]{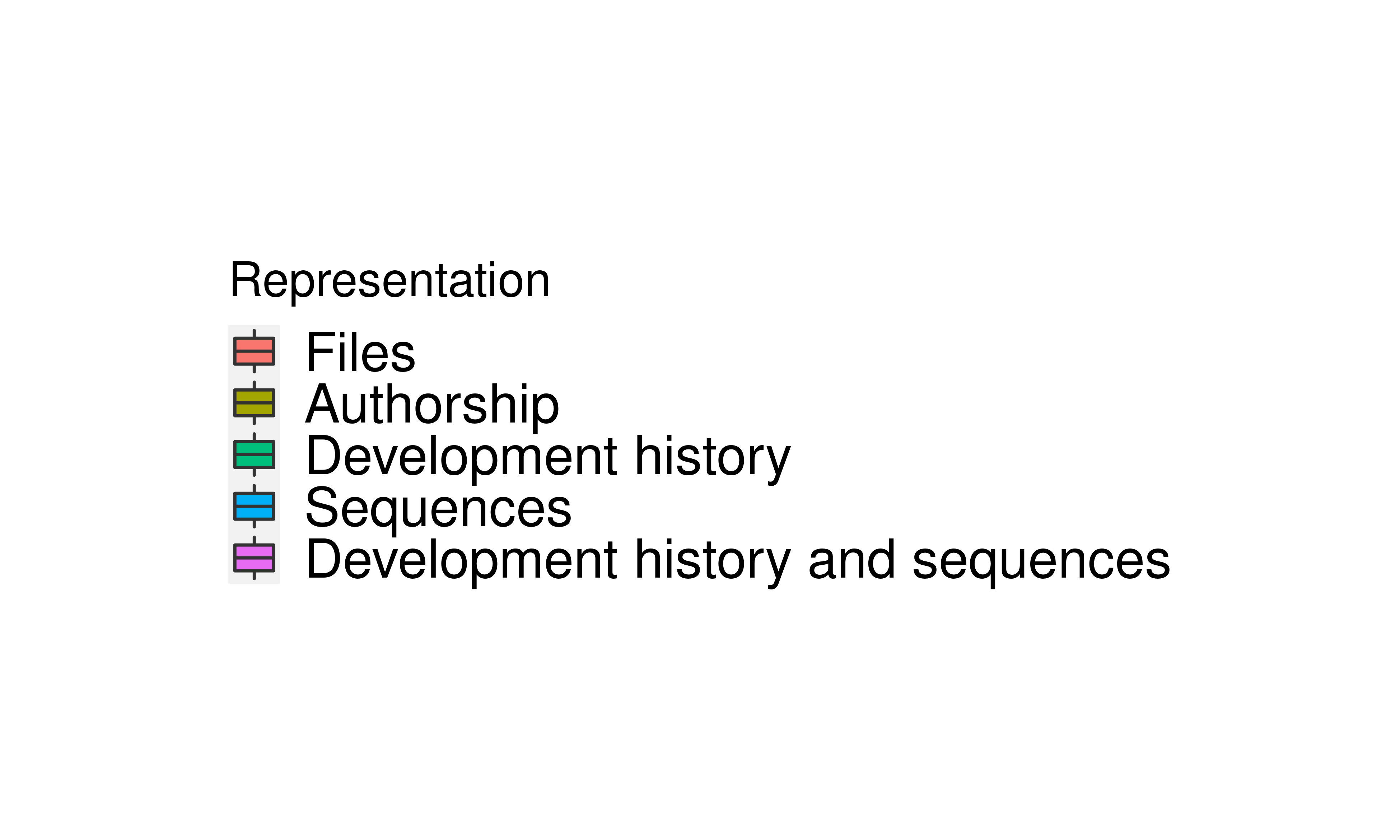} }

}

\caption{Grouping the decompositions by codebase and by number of clusters, and selecting the best decomposition in each group and each metric, highlights a difference between the representations and provides a different perspective on the results.}\label{fig:best-decompositions-compare}
\end{figure*}

Visually comparing the boxplots of the best decompositions of each
codebase, for each metric and for each monolith representation (Figure
\ref{fig:best-decompositions-compare}) shows that decompositions made
with the development history and the sequences of accesses
representations, simultaneously, display the best values in all metrics.
Statistical tests allow us to state with confidence that this is true
for uniform complexity, cohesion, coupling, and combined. This goes
against the conclusions drawn when considering all decompositions,
confirming our previous suggestion. For the case of \emph{tsr}, we
cannot say with confidence that this representation is better than just
the development history or the authorship, highlighting the poorer
performance that the sequence of accesses has in this metric.

\hypertarget{large-vs-small-codebases}{%
\subsection{Large vs small codebases}\label{large-vs-small-codebases}}

\begin{figure*}

{\centering \subfloat[Uniform complexity\label{fig:large-small-codebases-compare-1}]{\includegraphics[width=0.46\linewidth]{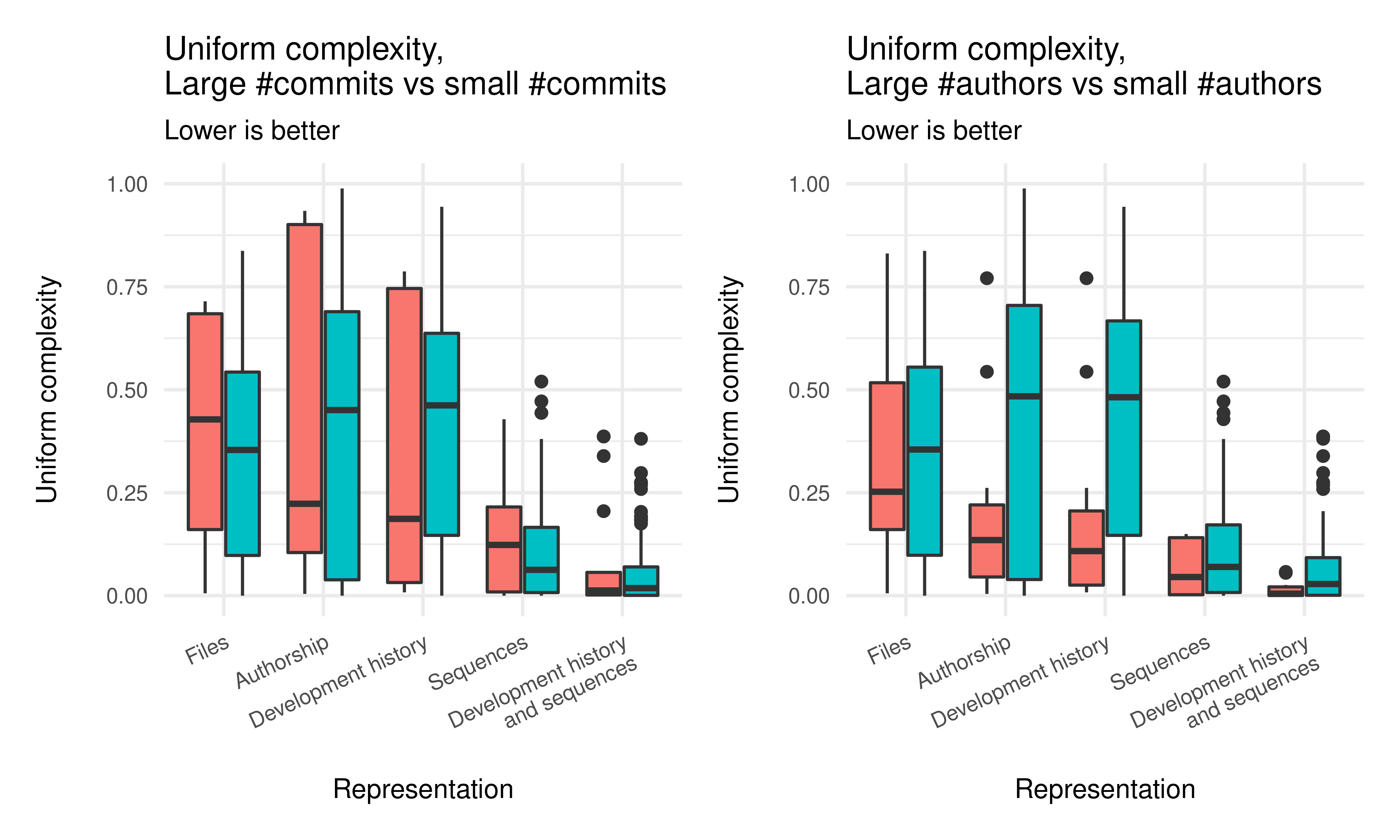} }\subfloat[Cohesion\label{fig:large-small-codebases-compare-2}]{\includegraphics[width=0.46\linewidth]{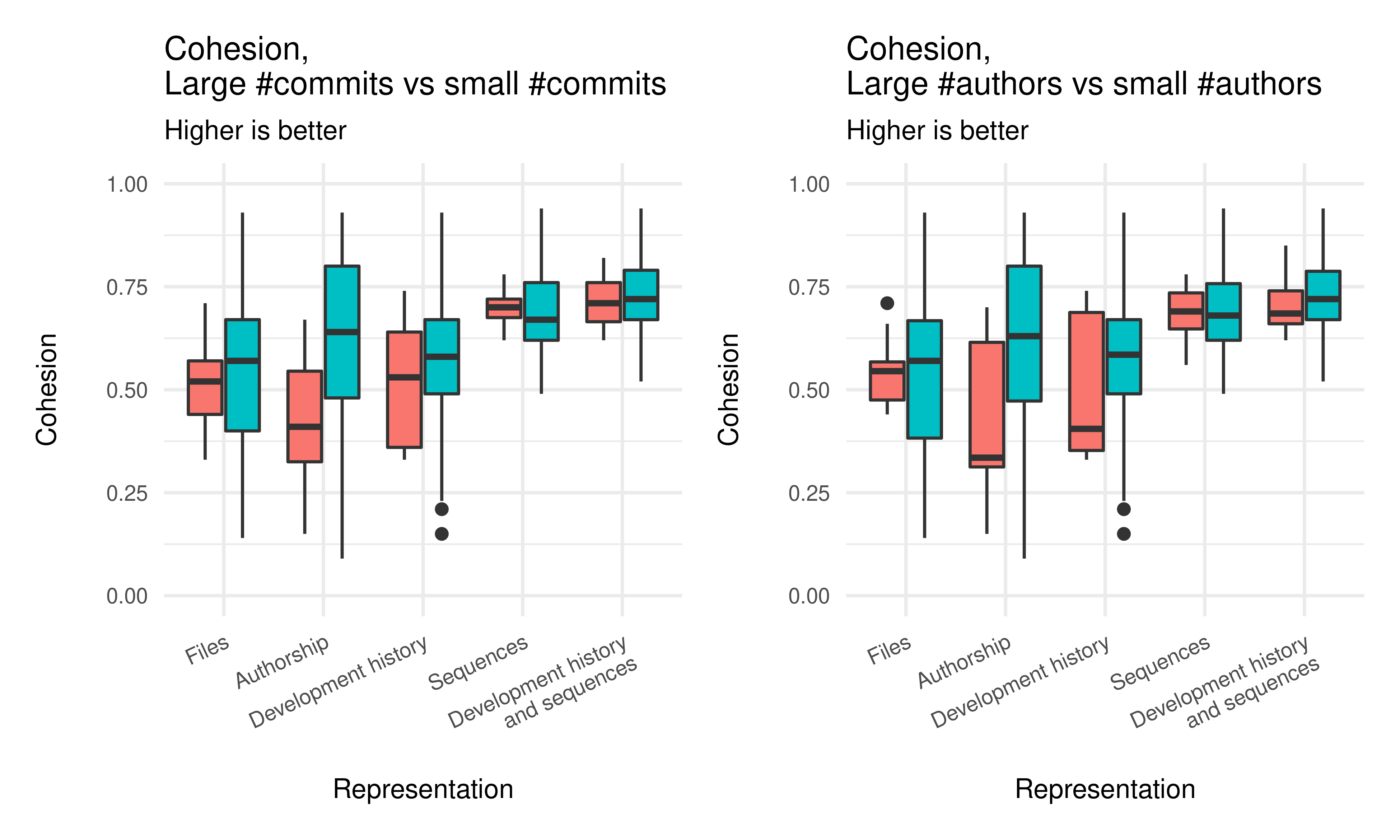} }\newline\subfloat[Coupling\label{fig:large-small-codebases-compare-3}]{\includegraphics[width=0.46\linewidth]{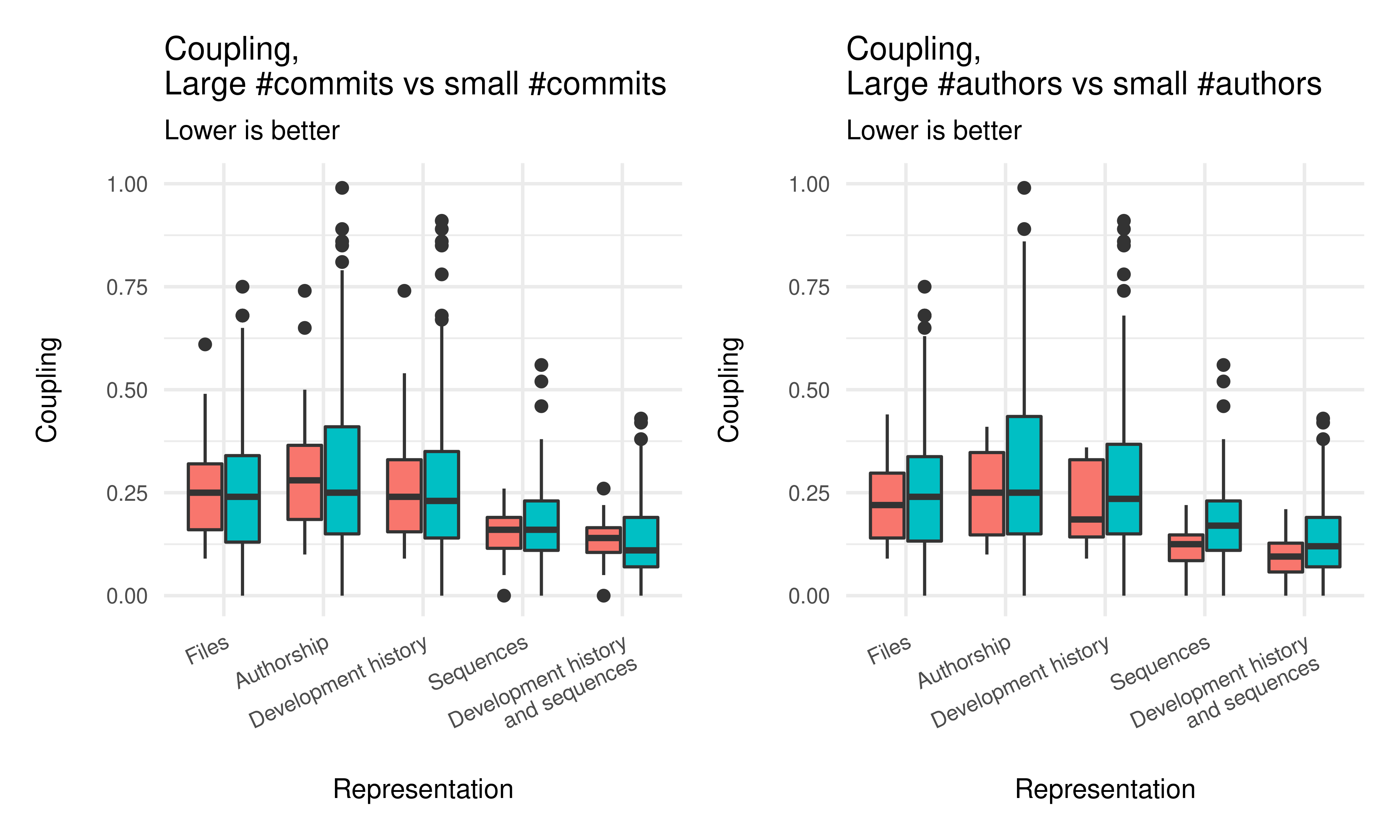} }\subfloat[tsr\label{fig:large-small-codebases-compare-4}]{\includegraphics[width=0.46\linewidth]{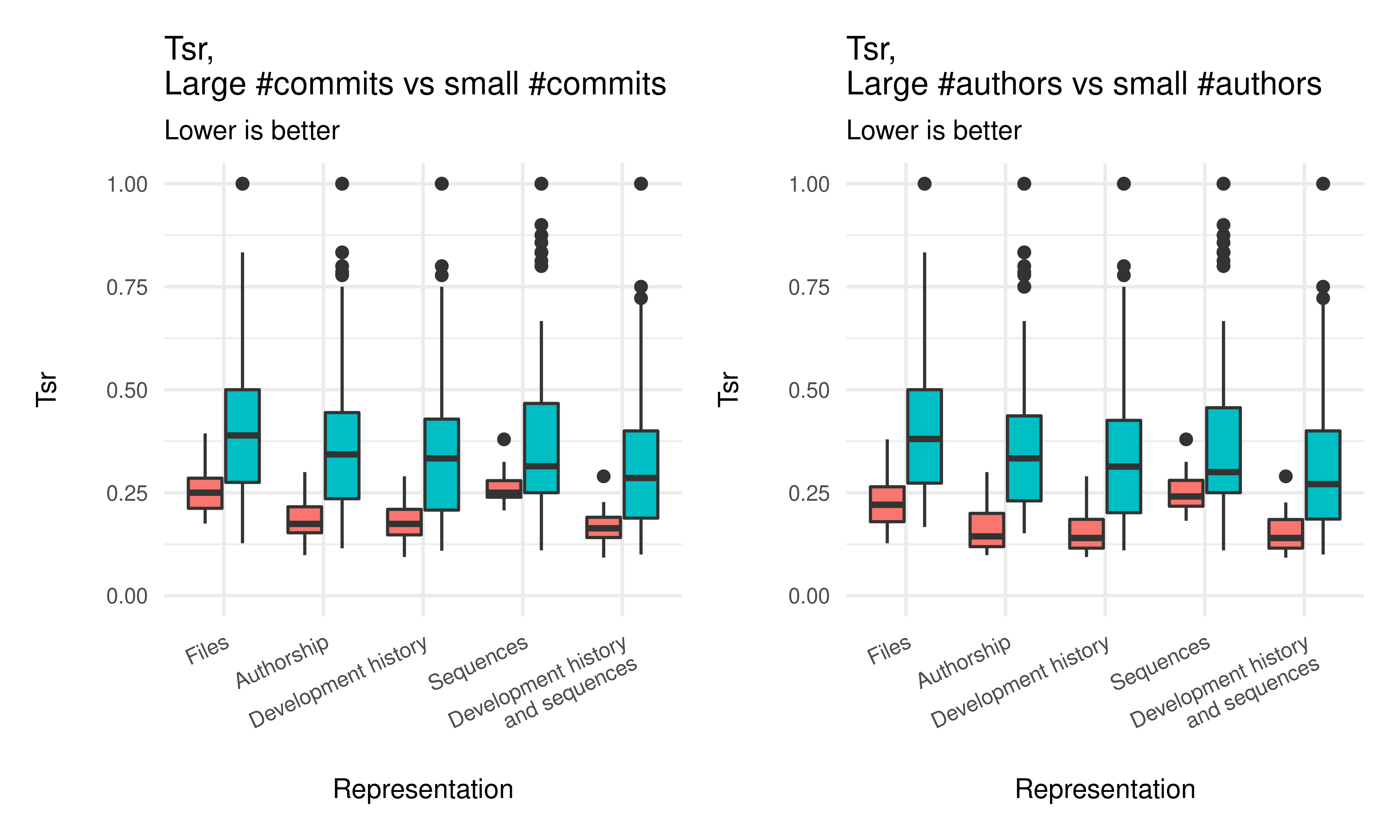}}\newline\subfloat[Combined\label{fig:large-small-codebases-compare-5}]{\includegraphics[width=0.46\linewidth]{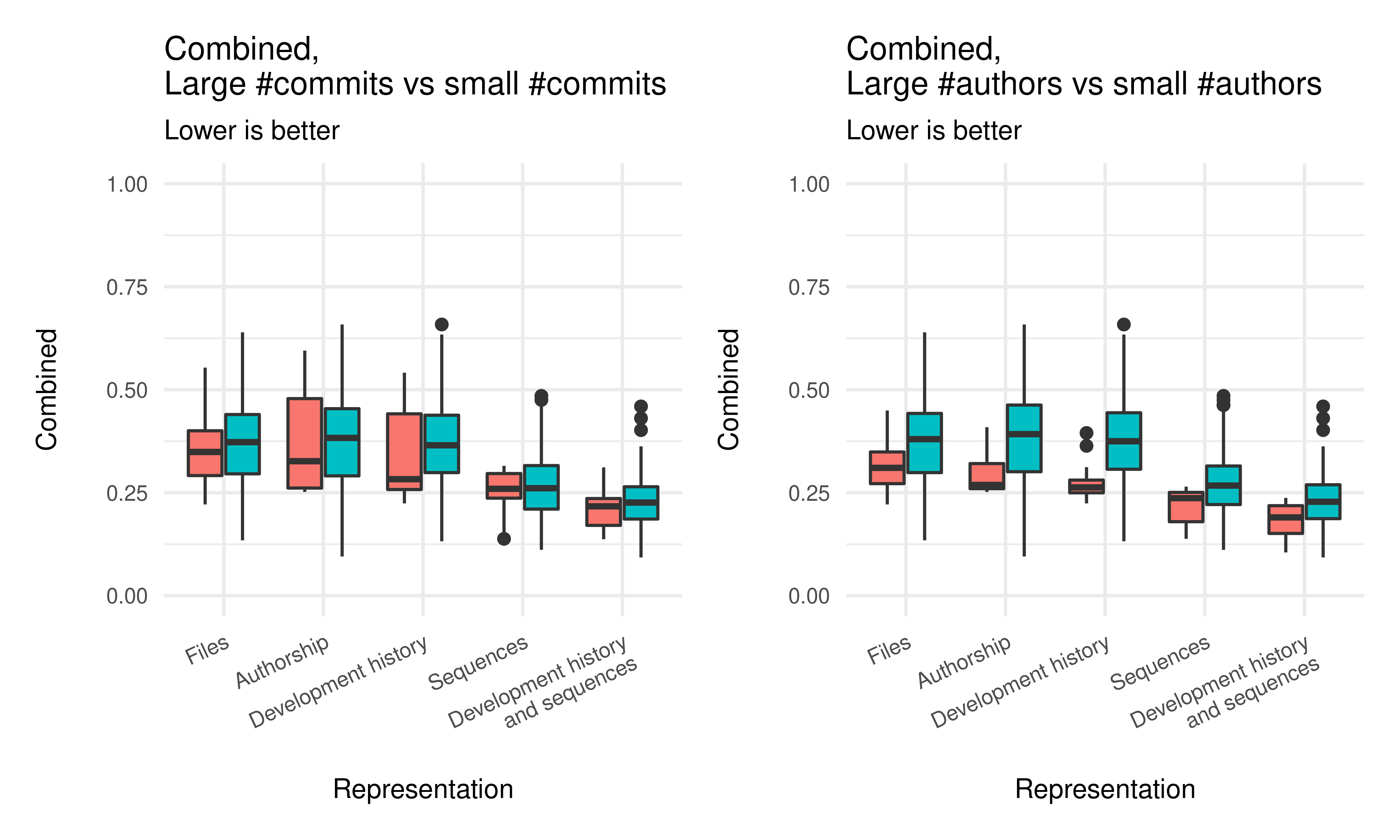} }\subfloat[Legend\label{fig:large-small-codebases-compare-6}]{\includegraphics[width=0.46\linewidth]{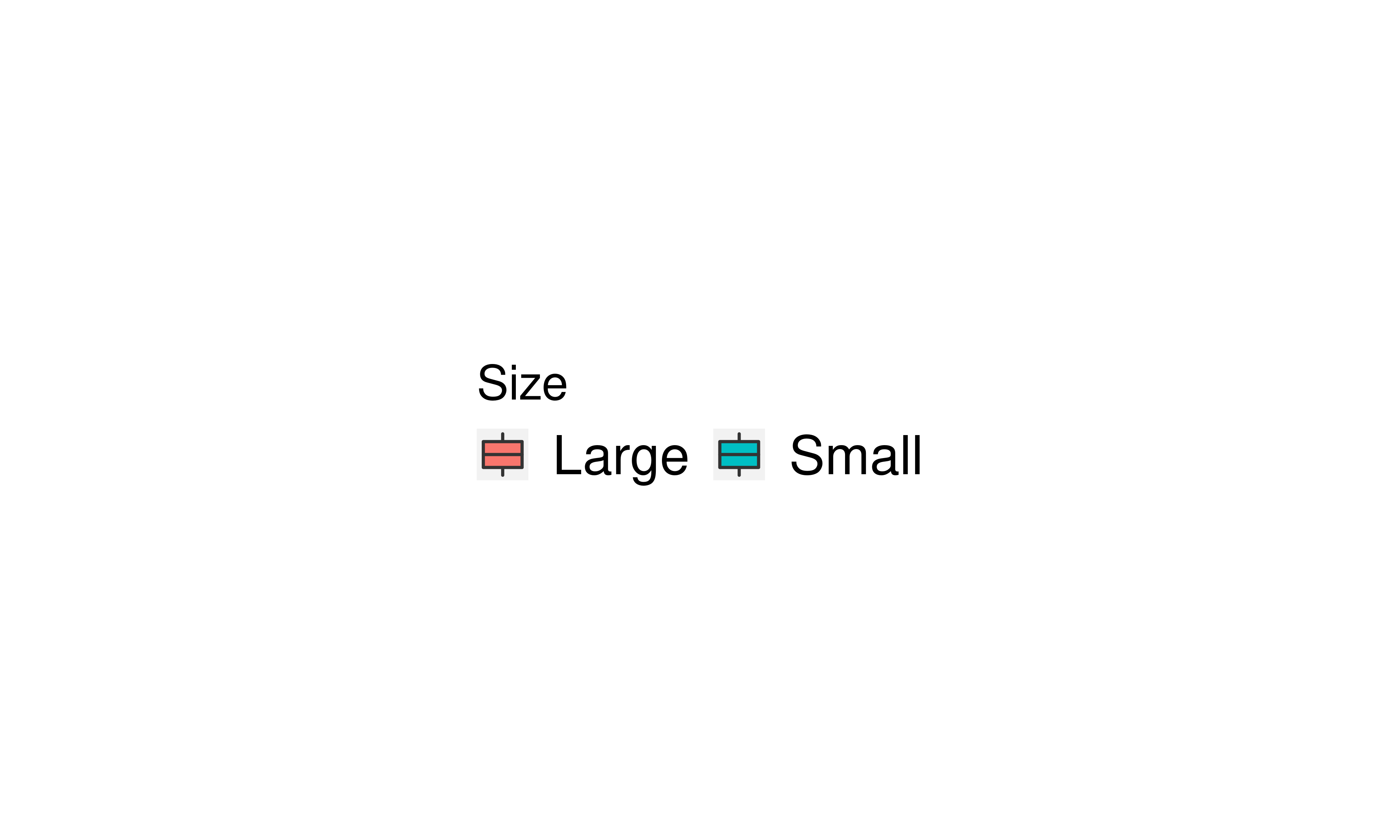} }

}

\caption{Comparison between the best decompositions of each representation. Different results are obtained when we compare large with small codebases.}\label{fig:large-small-codebases-compare}
\end{figure*}

In our study so far, we made no distinction between small and large
codebases. But, there are 3 codebases with more commits than the mean
plus one standard deviation, which we call ``large in the number of
commits''. The remaining codebases are called ``small in the number of
commits''. A similar classification is made for the number of authors of
each codebase. Considering the increased information available from the
development history in these large codebases, we asked ourselves if
their development history based representations perform differently than
the representations of small codebases. To test this hypothesis, we
gathered the best decompositions of each representation for each metric,
and compare the best decompositions of large codebases in the number of
commits with small codebases in the number of commits, as well as large
codebases in the number of authors with small codebases in the number of
authors. Box plots with these comparisons can be found in Figure~\ref{fig:large-small-codebases-compare}.

We can say with statistical significance that the median uniform
complexity of small codebases in the number of authors is greater than
that of large codebases in the number of authors, when comparing the
authorship representation for both cases (p-value: 0). We can also state
that there are no significant differences between the median uniform
complexity of the authorship representation in large codebases to the
median uniform complexity of the sequences representation in small
codebases, whereas there is a difference if we compare these two
representations in small codebases. This is a surprising result, as it
suggests the effectiveness of the authorship representation, when
compared with sequences, if more authors are present.

We can also say that the \emph{tsr} of the smaller codebases in the
number of authors is greater than the large ones in all representations.
Statistical tests reveal that there are no significant differences
between the authorship, development history, and the combined
development history and sequences representations, highlighting the
effect that the changes authorship data has in these representations.
Similarly to the uniform complexity, the development history based
representations of large codebases perform better than the sequence of
accesses of small codebases, whereas they perform worse in the case of
smaller codebases. Regarding coupling, the obtained values in the files
and authorship representations are very similar for both categories of
codebases, visually and with statistical significance, but the smaller
codebases display worse values in the development history, sequences,
and the development history and sequences representations. The good
performance of the larger codebases in the number of authors explains
the improvement of the combined metric, as the representations of
smaller codebases have statistically significant greater median values
than the large codebases. The only metric where the large codebases in
the number of authors do not display better results than the smaller
codebases is cohesion.

Large codebases in the number of commits display improved \emph{tsr}
median values across all representations, when compared with small
codebases in the number of commits. In the case of uniform complexity
and, consequently, combined, better results are obtained for the files
representation of large codebases (when compared with the files
representation of small codebases) only if we discard one of the large
codebases (Irida) that does not behave like the others. We find that the
modularity is neither improved nor worsened when we evaluate codebases
with more commits. In the case of cohesion, the smaller codebases in the
number of commits are similar to the large, and other than the
authorship representation, we do not find statistically significant
differences between their medians. For coupling, cannot confidently
state that the larger codebases have greater median than the smaller.
For the combined metric, we cannot state that the smaller codebases
display larger values in any representations, which means that overall,
the quality of codebases with more commits is comparable to the quality
of codebases with less commits.

Finally, we can state that just like in the previous analysis of the
best decompositions, the best results are obtained when combining the
development history with sequences of accesses, in all quality metrics -
both for large and small codebases in the number of commits and authors.





\subsection{Discussion}

We were able to properly compare the various representations in previous
sections, through visual analysis and statistical analysis. The median
quality metrics values across all codebases and clusters provided us
with an overview of the most common scenario. But this is a limited
view, as it does not fully represent what happens if an ideal
decomposition is found. To fix that, we also analyzed the best
decompositions of each metric and representation, where often different
results were obtained. Additionally, we often could not find significant
differences between the development history based representations, so we
separated the best decompositions according to the number of commits and
authors of their codebases, which gave us better insights in some
situations.

This whole analysis provided us with several conclusions that allow us
to answer the research question, focused around how the
sequences of accesses representation compares with development history
based representations:

\begin{itemize}
\item
  On average, the sequence of accesses representation is better in the
  case of complexity and coupling. For cohesion, no significant
  differences were found between representations, with the exception of
  the changes authorship representation, which performs worse than all
  others. On the other hand, representations with change authorship
  data, like the changes authorship, development history, and
  development history and sequences, perform better than the sequence of
  accesses for \textit{tsr}, although we have to highlight the still
  acceptable performance of the latter. These trade-offs are captured by
  the combined metric and confirmed through statistical tests, as we
  cannot state that any representation is better than the others at any
  number of clusters.
\item
  When looking at the best decompositions, the combined development
  history and sequence of accesses representation yielded the best
  values for all metrics. A notable exception is in the \textit{tsr} metric,
  where we could not state with statistical significance that this
  representation is better than the development history representation.
  Nevertheless, these conclusions give support to the notion that the
  presence of more data from different sources improves results.
  However, considering that this does not happen when looking at the
  vast majority of decompositions, it means that obtaining good results
  when using data from the development history is very dependent on
  choosing the ideal weights when creating the decomposition.
\item
  Through the comparison between large and small codebases in the number
  of commits and/or authors, we cannot state that the complexity of the
  changes authorship representation of large codebases in the number of
  authors is higher than the sequences of accesses representation of
  small codebases in the number of authors. Regarding cohesion and
  coupling, we found that the presence of more commits or authors does
  not improve results. On the other hand, more commits or authors does
  significantly improve the \textit{tsr} values, as any representation of
  larger codebases has a better median than the sequences of accesses
  representation of smaller codebases.
\end{itemize}

\hypertarget{threats-to-validity}{%
\subsection{Threats to validity}\label{threats-to-validity}}

Out of all decompositions, we find that \(0.37\%\) were made exclusively
with data from the development history, and \(9.52\%\) exclusively with
data from the sequences of accesses. The remaining \(90.11\%\) were made
with combined data. This is a consequence of using four measures related
to the sequences of accesses, but only two related to the development
history. To ensure this does not affect our findings, we opted to use a
statistical test that performs well even comparing groups with different
sample sizes, and do not rely only on boxplots to draw conclusions.
Additionally, note that these differences only applied for the first
analysis, where we consider to all the decompositions, which was rather
inconclusive. For all the other analyses, the best decompositions were
chosen and so, we have only one decomposition per representation and
number of clusters.

Not all repositories have a clean and linear history, with some
presenting many branches, refactors, and merges. This affects the
detection and processing of deletes and renames, which makes development
history based decompositions less efficient. Nevertheless, we obtained
good results for the development history representations.

We found that sometimes, files presented an \texttt{ADD} or
\texttt{MODIFY} change event after a \texttt{DELETE} event. In some
situations, this means we could be considering two distinct files as the
same one, if they happened to have the same filename and one of them was
deleted before the other was added. However, in all cases we found, the
files still existed in the latest repository snapshot and did correspond
to the same file that was deleted. Considering that this situation is
unlikely, and the existence of a \texttt{DELETE} event after an
\texttt{ADD} or \texttt{MODIFY} change event usually occurs due to
merges, we opted to still consider these files in our analysis, and we
don't discard them.

Our data collection approach was to gather data about all \texttt{.java}
files across all commits, and then discard non domain entities files
only in the decomposition phase. An alternative would be to filter all
commits and select those where only domain entities were changed. Our
approach is richer, as we have more data available and are not deleting
potentially useful relationships between files.

We adapted the logical coupling and the contributor coupling measures
from \cite{mazlami_extraction_2017}, by considering a fraction rather
than an absolute value. This was made to facilitate the integration of
other measures without much experimentation on the ideal weights that
would be required if absolute values were considered.

\section{Conclusions}
\label{sec:conclusion}

As the development of a monolith system progresses, it tends to get more complex, and introducing new features and bug fixes becomes harder. An architecture based on microservices allows for better scaling, so a migration from a monolith to this architecture brings plenty of advantages. To help with the migration, various automated approaches with different strengths, inputs, and evaluation metrics have been proposed, but they were generally tested on a reduced number of codebases and there is a lack of research on the comparison of different approaches.

In this work, we evaluated a total of 468k decompositions of 28 codebases, and compared their quality according to 5 metrics when created with different monolith representations: file changes, changes authorship, development history (which combines the previous two), sequences of accesses, and a combined development history and sequences of accesses.

On average, according to our quality metrics, development history based representations are not better than a sequences of accesses representation.  With respect to the best decompositions according to each metric, the vast majority of them (over $80\%$) were generated with the combined development history and sequences of accesses representation. This means that even if this combination does not produce the best results on average, when compared with other representations, it is very likely that the best weights configuration for a given metric considers both representations. Interestingly, we also found that codebases with a high number of authors present better decompositions, with the authorship representation of larger codebases achieving comparable results to the sequences of accesses of smaller codebases. On the other hand, an increased number of commits achieves similar results to a reduced number.

The code is publicly available\footnote{\href{https://github.com/socialsoftware/mono2micro/tree/joao-commit-complexity}{https://github.com/socialsoftware/mono2micro/tree/joao-commit-complexity}}, and a reproducible evaluation package is also provided\footnote{\href{https://github.com/socialsoftware/mono2micro/tree/joao-commit-complexity/data/commit/reproducible-evaluation}{https://github.com/socialsoftware/mono2micro/tree/joao-commit-complexity/data/commit/reproducible-evaluation}}.

\section*{Acknowledgment}
This work was partially supported by Fundação para a Ciência e Tecnologia (FCT) through projects UIDB/50021/2020 (INESC-ID) and PTDC/CCI-COM/2156/2021 (DACOMICO).

\bibliographystyle{./bibliography/IEEEtran}
\bibliography{./bibliography/biblio}

\end{document}